\documentstyle[psfig,11pt,aaspp4]{article}

\lefthead{Kuulkers et al.}
\righthead{GRB990510}

\begin{document}

%\title{BeppoSAX observations of the GRB990510 counterpart: a possibly
%beamed X-ray afterglow}

\title{GRB990510: on the possibility of a beamed X-ray afterglow}

\author{E.~Kuulkers$^{1,2}$, L.A.~Antonelli$^{3,4}$, L.~Kuiper$^1$, J.S.~Kaastra$^1$, 
L.~Amati$^5$, E.~Costa$^6$, F.~Frontera$^{5,7}$, J.~Heise$^1$, J.J.M.~in 't Zand$^1$, 
N.~Masetti$^5$, L.~Nicastro$^8$, E.~Pian$^5$, L.~Piro$^6$,
P.~Soffitta$^6$}

~\\

\affil{
$^1$ Space Research Organization Netherlands (SRON), Sorbonnelaan 2, 3584 CA Utrecht, The Netherlands\\
$^2$ Astronomical Institute, Utrecht University, P.O.\ Box 80000, 3508 TA Utrecht, The Netherlands\\
$^3$ BeppoSAX Scientific Data Center, Via Corcolle 19, I-00131 Roma, Italy\\
$^4$ Osservatorio Astronomico di Roma, Via dell'Osservatorio, I-00044 Monteporzio Catone, Italy\\
$^5$ Istituto Tecnologie e Studio Radiazioni Extraterrestri, CNR, Via Gobetti 101, I-40129 Bologna, Italy\\
$^6$ Istituto di Astrofisica Spaziale, CNR, Via Fosso del Cavaliere, I-00133 Roma, Italy\\
$^7$ Dipartimento di Fisica, Universit\`a di Ferrara, Via Paradiso 11, I-44100 Ferrara, Italy\\
$^8$ Istituto Fisica Cosmica e Applicazioni all'Informatica (CNR), via Ugo La Malfa 153, 90146 Palermo, Italy
}

\begin{abstract}

We discuss the prompt emission of the $\gamma$-ray burst (GRB) 990510
and its subsequent X-ray afterglow from 8.0 to 44.3\,hrs after the prompt emission, using
observations with the {\it Gamma-ray Burst Monitor} and {\it Narrow Field Instruments} on
{\it BeppoSAX}.  
In the 40--700\,keV band, GRB990510 had a fluence of
$\sim$1.9\,$\times 10^{-5}$\,erg\,cm$^{-2}$, whereas it reached a peak flux of
$\sim$2.4\,$\times$\,10$^{-6}$\,erg\,cm$^{-2}$\,s$^{-1}$.
The X-ray afterglow decay light curve can be
satisfactorily described by a single power law with index of
$-1.42 \pm 0.07$. 
Both the X-ray and optical behaviour of the afterglow 
can be explained by $\gamma$-ray burst debris expanding as a jet;
we find that the cooling frequency is (fixed) 
between the optical and X-ray wavelength bands.

\end{abstract}


\newpage 

\section{Introduction}

Among the about 25 $\gamma$-ray
bursts localized by the BeppoSAX {\it Wide Field Cameras} (WFCs), most
of those followed-up with the {\em Narrow Field Instruments} (NFIs)
onboard the same satellite have exhibited afterglows at X-ray energies
(e.g., Costa et al.\ 1997),
whereas less than half of them have exhibited afterglows in the
optical, IR, and/or radio (e.g., van Paradijs et al.\ 1997; Frail et
al.\ 1997).  Most X-ray afterglows show a smooth power-law decay
(with indices between $-$1.1 to $-$1.9),
the exceptions being GRB970508 (Piro
et al.\ 1998)  and GRB970828 (Yoshida et al.\ 1999), which exhibit
re-bursting events on time scales of a few hours and a day,
respectively, superimposed on a power-law trend. The brightest X-ray
afterglow so far, i.e.\ that of GRB990123, provided the first detection
of hard X-ray (15--60\,keV)  afterglow emission (Heise et al.\ 2000). 

Here we discuss BeppoSAX observations of the prompt $\gamma$-ray emission and the 
X-ray afterglow of GRB990510.
On 1999 May 10 the BATSE experiment onboard the {\it Compton Gamma Ray
Observatory} (CGRO) was triggered by GRB990510 at 8:49:06.29 UT (trigger
7560, see Kippen et al.\ 1999). The GRB was also detected by the
BeppoSAX {\it Gamma-Ray Burst Monitor} (GRBM; Amati et al.\ 1999a) and WFC
unit 2 (Dadina et al.\ 1999; Briggs et al.\ 2000), 
as well as by {\it Ulysses} (Hurley et al.\ 2000) 
and the {\it Near Earth Asteroid Rendezvous} (NEAR) spacecraft
(Hurley 1999, private communication).  In the WFC energy range (2--28\,keV)  the GRB had a
duration of $\sim$80\,s and reached a peak intensity of 4.3~Crab. The
WFC
error box was followed up in X-rays by the Narrow Field Instruments (NFIs)
onboard BeppoSAX $\sim$8\,hrs after the event and a strong decaying source
was found (Piro et al.\ 1999b, Kuulkers et al.\ 1999). 
About 8.5\,hr after the $\gamma$-ray/X-ray event the optical counterpart
was found (Vreeswijk et al.\ 1999a) with a redshift of $z>1.62$ (Vreeswijk
et al.\ 1999b). A linear polarization of 2\%\ was measured
(Covino et al.\ 1999; Wijers et al.\ 1999). 
Extended emission around the optical counterpart of GRB990510 has not been
clearly detected, which indicates that a possible underlying host galaxy
must be very faint (Israel et al.\ 1999; Fruchter et al.\ 1999b; Beuermann
et al.\ 1999).

The light curve of the optical afterglow of GRB990510 does not follow a simple power-law
decay, but showed smooth steepening after about one and a half day after the $\gamma$-ray burst
(Harrison et al.\ 1999; Stanek et al.\ 1999; Israel et al.\ 1999). 
Traces of such a characteristic have also been found in the optical
afterglow of GRB990123 (Kulkarni et al.\ 1999; Fruchter et al.\ 1999a) and the
near-infrared afterglow of GRB990705 (Masetti et al.\ 2000).
It has been regarded as the signature of a  
decreasing collimation in a relativistic flow
(Sari et al.\ 1999; Rhoads 1999).  
Such behavior has never been
observed in the X-ray afterglows of GRBs.  The relatively
large brightness of the GRB990510 X-ray afterglow allows an excellent 
opportunity to study the X-ray light curve in search of such a feature. 

\section{Observations}

\subsection{GRBM}

The GRBM consists of the 4 anti-coincidence shields of the {\it Phoswich
Detection System} (PDS; Frontera et al.\ 1997; Costa et al.\
1998).  The GRBM detector operates in the 40--700\,keV energy band. The
normal directions of two GRBM shields are co-aligned with the pointing
direction of the WFCs. The on-axis effective area of the
GRBM shields, averaged over the 40--700\,keV band, is 420\,cm$^2$.
The data from the GRBM include rates with
1\,s time resolution and energy ranges of 40--700\,keV and $>$100\,keV,
and average 240-channel spectra in the 40--700\,keV band every 128\,s
(independently phased from GRB trigger times).  
For our spectral analysis we use data in the 70--650\,keV
band, since in this energy range the GRBM 240-channels response matrix is
known with sufficient accuracy. For studying the GRB
spectral evolution we use the 1\,s ratemeters, and we check their
consistency with the GRB time averaged spectra obtained from the 240
channel data (see Amati et al.\ 1999b).  

GRB990510 was detected by the GRBM, but the instrument was not
triggered to a GRB data acquisition mode, because a previous false
event prevented this.  Therefore, no high time resolution data have
been acquired for this burst and the time resolution is limited to
1\,s. 

\subsection{NFI}

The NFI include two imaging and two non-imaging instruments.  The
imaging instruments are the {\it Low-Energy Concentrator Spectrometer}
(LECS), sensitive from 0.1 to 10\,keV (Parmar et al.\
1997), and the {\it Medium-Energy Concentrator Spectrometer} (MECS),
sensitive from 2 to 10\,keV (Boella et al.\ 1997). They
both have circular fields of view with diameters of 37$\arcmin$ and
56$\arcmin$, respectively.  The non-imaging instruments are the {\it
Phoswich Detector System} (PDS), sensitive from 13 to 300\,keV 
(Frontera et al.\ 1997), and the {\it Gas Scintillation
Proportional Counter}, sensitive from 4 to 120\,keV (Manzo
et al.\ 1997). In our analysis we used data from the LECS, MECS and
PDS. 

The 3$\arcmin$ radius WFC error box of GRB990510 was observed by the NFI
from 8.0 to 44.3 hours after the BATSE trigger time, i.e.\ from
MJD\,51308.70--51310.22 (UT 1999 May 10.70--12.22). The total LECS, MECS
and PDS on-source exposure times were 31.7, 67.9 and 41.5\,ksec,
respectively.

\section{Data analysis}

\subsection{Prompt $\gamma$-ray emission}

The GRBM light curve of the burst is shown in Fig.~1 (top). 
Two main pulses $\sim$40\,s apart are observed. Between these pulses
the GRB flux level is consistent with zero. The first pulse contains
two sub-pulses with peak fluxes in the ratio 3:1.  The second pulse
consists of 5 sub-pulses with the first two having the highest peak
flux and the following three being much weaker (by a factor of about
6).  The entire GRB duration is 75\,s. The GRB fluence in the
40--700\,keV band is (1.9$\pm$0.2) $\times 10^{-5}$\,erg\,cm$^{-2}$,
while the peak flux reached in the same energy band is (2.4\,$\pm$0.2)
$\times 10^{-6}$\,erg\,cm$^{-2}$\,s$^{-1}$ (all errors quoted in this
paper are 1$\sigma$ uncertainties, unless noted otherwise.) 

We performed a spectral analysis of the prompt emission of GRB990510 in
the 70--650\,keV band, by following approximately the same procedure 
used for, e.g., GRB970228 and GRB980329 (see Frontera et al.\ 1998; In 't
Zand et al.\ 1998).
The average spectrum of the prompt emission of GRB990510 can be
satisfactorily described ($\chi^2$ = 6.84 for 9 d.o.f.) by a broken 
power-law, with a break energy $E_{\rm break}=200\pm27$\,keV and 
power-law indices before and after the break of $-1.36\pm 0.16$ and
$-2.34\pm 0.24$, respectively.
We note that a fit to 
the canonical $\gamma$-ray burst spectral 
model as introduced by Band et
al.\ (1993) is also acceptable ($\chi^2$ = 7.65 for 9 dof); however, 
we could not constrain the value of the 
power-law index below the peak energy $E_p$, $\alpha_B$
($\alpha_B=-0.7\pm 0.8$).
The other Band parameters, i.e.\ the break energy or cut-off energy, $E_0$, 
and power-law index above $E_p$, $\beta_{\rm B}$, in this fit 
are $184\pm 76$\,keV and $-2.68\pm 0.64$, respectively.

We also studied the spectral evolution of the prompt emission 
by assuming that a power-law $F(E) \propto E^{-\Gamma}$ connects the two
energy bands 40--100\,keV and 100--700\,keV, and that the burst flux above
700\,keV is negligible.  The photon index $\Gamma$, computed as a color
index between the two energy ranges, is reported in the bottom panel of
Fig.~1: the spectrum seems to slightly soften during the first main pulse.
When the second main pulse starts, the spectrum is harder before it
softens. The indices have typical values for GRBs.

\subsection{X-ray afterglow}

The combined image from MECS units 2 and 3 
shows clearly the presence of a previously unknown bright source within
the WFC error circle, formerly proposed as the X-ray counterpart of the
GRB (Piro et al.\ 1999b, Kuulkers et al.\ 1999). It is elongated toward
the NNW direction (see Fig.~2). This extension is likely due to the
presence of an unresolved point source, partially contaminating the
bright source. Another X-ray source is present $\sim$13\,arcmin NNW of the bright
source, and outside the WFC error circle of GRB990510.  The LECS image, albeit less
exposed than the MECS image, also shows a bright source, with a similar extension.

The proximity of the probable X-ray afterglow candidate to the weaker,
contaminating source makes an accurate spatial analysis necessary,
accounting for the extended tails of the point-spread functions of the
LECS and MECS instruments, in order to separate the two point sources. We
used different independent approaches to resolve this problem and
performed fits of the resulting spectra.
We here discuss one of these approaches, 
the maximum likelihood method, since the main results of this paper
are obtained using this approach. The other two approaches serve 
as a consistency check of the former approach and are described in 
Appendix~A.
For all spectral fits, the background was evaluated from blank sky
observations, after checking its stability with several source-free
regions around the X-ray afterglow.  The LECS spectrum was considered only
in the 0.2--4\,keV interval, due to calibration uncertainties above
$\sim$4\,keV. The MECS spectrum was considered in the 2--10\,keV range.

In the maximum likelihood method one searches for single point sources on top of a
background model (assumed to be flat, i.e.\ cosmic diffuse X-ray
and particle-induced background).  With this method, which allows a
simultaneous analysis of several sources, one retrieves all photons from
the point sources as detected by the instrument (for a more detailed
description of the method we refer to Kuiper et al.\ 1998 and In 't Zand et al.\ 2000). 

This method shows that the LECS and MECS image can be
satisfactorily described by three point sources on top of a flat
background model.  Their best-fit positions from the MECS 
measurements are given in Table~1, together
with the source designations as given by Kuulkers et al.\ (1999).  In
Fig.~2, we show the maximum likelihood map using the MECS data
(2--10\,keV)  sampled over the full duration of the observation.  In this
figure we also show the best derived WFC position of the prompt emission
(Dadina et al.\ 1999), the {\it Ulysses}/GRBM triangulation annulus of the
prompt emission (Hurley et al.\ 2000) and the position of the optical
afterglow (Vreeswijk et al.\ 1999a).  Since 1) 1SAX\,J1338.1$-$8030
is positionally consistent with the position of GRB990510 as derived by
the WFC and the triangulation annulus, 2) the optical afterglow of
GRB990510 lies within the confidence contours of this X-ray source, and 3) 
the X-ray emission of 1SAX\,J1338.1$-$8030 decayed during our observations
(see below) we conclude that 1SAX\,J1338.1$-$8030 is the X-ray
afterglow of GRB990510. 

We note that no other source near 1SAX\,J1337.6$-$8027
has been reported previously at other wavelengths, while
1SAX\,J1336.0$-$8018 lies close (within $\sim$2$\arcmin$) to the radio
source PMN\,J1335$-$8016 (Wright et al.\ 1994). 

X-ray spectra of 1SAX\,J1338.1$-$8030 were
generated in 9 energy bins, logarithmically distributed between
0.2 and 4.0\,keV for the LECS data, and in 20 energy bands, 
logarithmically distributed between 1.6 and 10\,keV for the MECS data.
The best-fit value of the normalization between the LECS and MECS was found to
be $\sim$0.7.  This is within the range usually found (0.6--0.9, see e.g.,
Favata et al.\ 1997; Piro et al.\ 1999a).  We therefore fixed this
normalization to 0.7. 
For 1SAX\,J1337.6$-$8027 we performed spectral fits only using the MECS data
and fixed the hydrogen column density, N$_{\rm H}$, to that found 
for 1SAX\,J1338.1$-$8030.
The best-fit parameters of the mean
spectra using the maximum likelihood method, and that of the other
methods, are given in Table~2.
We note that the best-fit values of N$_{\rm H}$
are close to that
derived for the mean Galactic value from the H{\sc I} maps by Dickey \&\
Lockman (1990) in the region of GRB990510, i.e.\
0.94\,$\times$\,10$^{21}$\,atoms\,cm$^{-2}$. 
We derive an unabsorbed average flux of $\sim$1.47\,$\times$\,10$^{-12}$\,erg\,s$^{-1}$\,cm$^{-1}$
(2--10\,keV) for 1SAX\,J1338.1$-$8030 during our observation.

To search for possible changes of the afterglow spectral shape 
during the observation, we logarithmically divided the whole
MECS observation into three time bins so that in each time interval there were
approximately equal amounts of counts.  The spectra are at all times
statistically well described by a single power-law, whose index does not
change significantly during the afterglow decay (see Table~3).

It has been suggested that (red-shifted) iron K lines may 
be present in the spectra of X-ray afterglows (Piro et al.\ 1999a; Yoshida et
al.\ 1999).  Since the redshift to GRB990510 has been reported to be
$>$1.62 (Vreeswijk et al.\ 1999b), one might expect such a line below
2.5\,keV.  We do not see, however, clear evidence for lines in this
region, neither in the total spectrum nor in the three time intervals,
with 90\%\ confidence upper limits on the line intensity of typically
7\,$\times$\,10$^{-6}$\,photons\,s$^{-1}$\,cm$^{-1}$ for the total
averaged afterglow spectrum.

The X-ray afterglow of GRB990510 was not detected with the PDS
instrument. By assuming a power-law spectrum with a photon index of
$-$2.1, the 2$\sigma$ upper limits on the flux from the GRB990510
region are 2.6\,$\times$\,10$^{-12}$\,erg\,s$^{-1}$\,cm$^{-1}$ and
5.0\,$\times$\,10$^{-12}$\,erg\,s$^{-1}$\,cm$^{-1}$, for the energy ranges
15--30\,keV and 15--60\,keV, respectively.  This is consistent with that
estimated from extrapolation of the LECS/MECS spectra.

We obtained light curves in the MECS 2--10 keV range of the
individual sources in 15 temporal bins, logarithmically spaced in time
since the BATSE trigger (Fig.~3).  
1SAX\,J1338.1$-$8030 clearly fades during our observations. By fitting
this decline with a power-law $I$(t)\,$\propto ({\rm t}-{\rm
t}_0)^{\alpha_X}$, we obtain $\alpha_X = -1.42 \pm 0.07$
($\chi^2$/dof=10.2/13). The corresponding fit is shown in Fig.~4 (solid
line). The count rate of 1SAX\,J1337.6$-$8027
is consistent with being constant ($\chi^2$=13.4 for 14 d.o.f.)  at
$\sim$0.003\,cts\,s$^{-1}$. 

\section{Discussion}

GRB990510 ranks among the top 25\%\ of the brightest GRB observed by the GRBM,
while it ranks among the top 4\%\ (9\%\/) of the BATSE burst flux (fluence) distribution
(Kippen et al.\ 1999). The mean prompt $\gamma$-ray spectrum is well
described by a broken power-law, with a break
energy of $\sim$200\,keV. The fluence, peak flux 
and spectrum as measured with the GRBM are comparable to
those measured with BATSE. With the repeated pulses and
"hard-to-soft'' spectral evolution, the $\gamma$-ray
light curve and spectral behavior of GRB990510 are reminiscent of
GRB970228 (Frontera et al.\ 1998).

The X-ray counterpart of GRB990510, 1SAX\,J1338.1$-$8030, is also very bright if
compared with other GRB X-ray afterglows (see e.g.\ Piro 2000), 
and decays according to a typical power-law
with index $-$1.42, which is consistent with that expected in relativistically
expanding fireball models (e.g., Wijers, Rees \&\ M\'esz\'aros 1997). 
However, the optical light curve smoothly steepens
$\sim$1--2 days after the prompt $\gamma$-ray emission
(Stanek et al.\ 1999; Harrison et al.\ 1999; Israel et al.\
1999; see also Fig.~4). It was found that this steepening occurs at the same
time in the different optical bands. 
To characterize its shape, the (V,R,I)-band data were
simultaneously fitted by Harrison et al.\ (1999) with the following 
four-parameter function\footnote{As noted by Harrison et al.\ (1999), the
function which describes the optical (B,V,R,I) light curve by Stanek et
al.\ (1999) and Israel et al.\ (1999) is different, leading to somewhat different values of the
break time, i.e.\ $\sim$1.57~days.}:

\begin{equation}
F_{\nu}(t)=f_{\star}(t/t_{\star})^{\alpha_1}[1-\exp(-J)]/J;\,\,\,\,\,J(t,t_{\star},\alpha_1,\alpha_2)=(t/t_{\star})^{(\alpha_1-\alpha_2)},
\end{equation}

with $t_{\star}=1.20\pm 0.08$\,days, $\alpha_1 = -0.82\pm
0.02$, and $\alpha_2 = -2.18\pm 0.05$. 
In Fig.~4 we plot the optical
R-band data taken in the same time span as the X-ray data, together
with the above described function.  It
is clear that the optical data are not consistent with a power-law
decay in that time span. 
We fitted the X-ray afterglow light curve with the same function
as above, while fixing the decay indices to
those derived in the optical. We find that the corresponding fits are
bad, with $\chi^2$ values of 24--27 for 14 d.o.f., depending on which
parameter values we use among those reported by the different authors 
(Stanek et al.\ 1999, Harrison et al.\ 1999, Israel et al.\ 1999). 

A steepening in the light curves can be expected in the fireball model
when the cooling frequency moves towards lower frequencies in the observed
frequency range. In that case the decay index $\alpha$ changes by 0.25
(Sari et al.\ 1998).
However, the steepening of the optical decay is independent of wavelength (or achromatic) 
and the optical decay index $\alpha$ changes by $\sim$$-$1.36 (Harrison et al.\ 1999).
We provide additional evidence against a changing cooling frequency.
In that case one would expect the optical decay index to be similar to that in the X-ray band, 
in contrast to what is observed.

It has recently been realized that not all afterglow light curves are
consistent with emission from expanding shells that are spherically
symmetric, but that beaming may be important
(i.e., jets, see e.g.\ Sari et al.\ 1999; Rhoads 1999). Such
jets explain the presence of the steepening observed in the
optical afterglow light curves of GRB990510 (e.g.\ Harrison et al.\
1999).  Sari et al.\ (1999) presented general expressions for the
expected spectral and decay index, appropriate for both spherical shell and
jet evolutions shortly after the $\gamma$-ray event.  Our observed
X-ray spectral index of $-$1.03$\pm$0.08 implies a value of the index $p$ of
the electron energy distribution in the expanding material of $p\simeq 2.1$ in
the case of fast cooling (i.e., when the cooling frequency is below the
X-ray range). In the alternate case (i.e., the cooling frequency is
above the X-ray range) we derive $p\simeq 3.1$.  Harrison et al.\
(1999) found that the optical light curves can only imply 
$p\simeq 2.1$, where the cooling frequency is above the optical wavelength
range. Therefore, we conclude that the cooling frequency is 
between the optical and X-ray wavelengths. 
Note that the cooling frequency stays constant for a spreading jet 
(Sari et al.\ 1999).

At early times after the burst the decay light curve of a
collimated source is identical to that of a spherical one, since then
only a small portion of the emitting surface is visible due to
relativistic beaming (the opening angle then is $1/\gamma$, where $\gamma$
is the Lorentz factor).  In that case the decay index, $\alpha$, is
expected to be $-(3p-2)/4\simeq -1.1$ 
in the case of fast cooling (i.e.\ steeper than in the optical: 
$-(3p-1)/4\simeq -1.3$; Sari et al.\ 1999).  As the fireball
evolves, $\gamma$ decreases, and the beaming angle will eventually
exceed the jet opening angle. At that time one will see a break in the
light curve, with $\alpha = p \simeq 2.1$, while the optical and X-ray
decay index are similar after the break.  Therefore, we fitted the X-ray
afterglow light curve again, now fixing $\alpha_1$ and $\alpha_2$ to
$-$1.1 and $-$2.1, respectively, and $t_{\star}$ to that found in the
optical. This leads to good fits with $\chi^2$ values of about 12 for
14 dof. The corresponding fit is also shown in Fig.~4 (dotted line)
with extrapolations to the boundaries of the plot.  This shows that the
observed X-ray afterglow of GRB990510 is consistent with the jet
interpretation.  As evident from Fig.~4, X-ray observations of the very
early afterglow or a long time after the break time could have
clearly discriminated whether the X-ray afterglow light curve is
described by a single power law or consistent with the jet
interpretation.

We conclude that, even if we could not distinguish the presence of a clear break in
the X-ray light curve, the only explanation within the fireball model
consistent with the X-ray 
and optical data is a jet evolution, where the cooling frequency
lies between the optical and X-ray wavelengths.
Future observations of afterglows at late times with the X-ray
observatories recently launched (Chandra and XMM-Newton) may provide a
direct evidence of such a temporal X-ray feature.

\acknowledgements

The BeppoSAX mission is a joint Italian and Dutch program.  We thank
M.R.~Daniele, S.~Rebecchi (SDC, Telespazio, Rome), G.~Scotti (SOC,
Telespazio, Rome) and G.~Gennaro (OCC, Telespazio, Rome), for their prompt
help in coordinating the ToO observations and the preparation of the FOT.
We made use of the SIMBAD astronomical database.

\appendix

\section{Other image analysis methods} 

\subsection{`SPEX' method}

One of the other methods to investigate source-rich regions 
is currently implemented in the X-ray spectral fitting code SPEX
(Kaastra et al.\ 1996). This approach consists of simultaneously
fitting the spectra from different detector sections, taking into account
the spill-over from photons coming from one detector section into another
detector section of the sky.  For a detailed description of this method
we refer to Vink et al.\ (2000) and Kaastra et al.\ (2000).  
The detector sections we used are two circles of 3$\arcmin$
radius (which limits the effect of contamination due to source proximity),
centered on the positions of 1SAX\,J1338.1$-$8030 and 1SAX\,J1337.6$-$8027
obtained from the maximum likelihood method. 
Note that in the case of the LECS a non-negligible amount of photons will
lie outside the extraction regions, especially at low energies
($\gtrsim$20\%\ for $\lesssim$1\,keV in the case of a 3$\arcmin$
extraction radius), due to its relatively large point spread function;
this leads to some degradation in the sensitivity at these energies.
The extraction of the spectra and the generation of response matrices
takes into account the characteristics of both the LECS and MECS.  The
spectral resolution is oversampled by the LECS and MECS energy channels,
therefore we rebinned the spectra so that roughly each resolved energy bin
contains three spectral channels.

The hydrogen absorption column density, N$_{\rm H}$, was forced to be
similar for both 1SAX\,J1338.1$-$8030 and 1SAX\,J1337.6$-$8027, since leaving
them both free led to unstable fits for 1SAX\,J1337.6$-$8027. The
best-fit value of the normalization between the LECS and MECS was 
found to be $\sim$0.8, and we therefore fixed it to 0.8. 
Both the spectrum of 1SAX\,J1338.1$-$8030 and
1SAX\,J1337.6$-$8027 are simultaneously well described by power-law models
(subject to interstellar absorption).  The best-fit parameters for both
sources are given in Table~2. 

\subsection{"Canonical'' method} 

We also evaluated the 
spectral fitting results offered by the maximum likelihood method,
by following, for the spectral extraction, the classical method, which is
more or less appropriate only for isolated point sources. 
We extracted the spectra within circles of radius 8$\arcmin$ and
4$\arcmin$, for the LECS and MECS, respectively, centered on the
best-fit position of 1SAX\,J1338.1$-$8030 obtained with the maximum-likelihood
method.  
The resulting spectra were also rebinned to bin sizes corresponding to
roughly one third of the detector spectral FWHM resolution, with the additional
constraint that each bin contained at least 20 counts. 

Since the fitted centroids of 1SAX\,J1338.1$-$8030 and of
1SAX\,J1337.6$-$8027 are only 3.8 arcmin apart, the resulting spectra
clearly contain the summed contribution of the two sources.  
Therefore, it is expected that the results of the
spectral fits are reasonable if the X-ray flux of 1SAX\,J1337.6$-$8027 is
negligible with respect to that of 1SAX\,J1338.1$-$8030 and/or if
1SAX\,J1337.6$-$8027 does not vary over the observed time interval. 
Since the maximum likelihood method showed that the spectrum
and emission level of 1SAX\,J1337.6$-$8027 do not significantly vary with
time, we fixed the contribution from 1SAX\,J1337.6$-$8027 to that found by
the SPEX method and fitted the afterglow spectrum by leaving the hydrogen
column density as a free parameter.  The best-fit parameters are also reported
in Table 2. 

\newpage

\begin{table}
\caption{Maximum likelihood positions (epoch J2000.0) for the three MECS sources}
\begin{tabular}{l|ccc}
\hline
Source & $\alpha$ & $\delta$ & r$_{\rm err}$$^a$\\ 
\hline
1SAX\,J1338.1$-$8030 & 13$^h$\,38$^m$\,11$\fs$22 & $-$80$\arcdeg$\,30$\arcmin$\,02$\farcs$1 & 2$\farcs$4 \\
=GRB990510's afterglow & & & \\
1SAX\,J1337.6$-$8027 & 13$^h$\,37$^m$\,46$\fs$46 & $-$80$\arcdeg$\,26$\arcmin$\,31$\farcs$3 & 7$\farcs$2 \\
1SAX\,J1336.0$-$8018 & 13$^h$\,36$^m$\,02$\fs$71 & $-$80$\arcdeg$\,18$\arcmin$\,22$\farcs$8 & 7$\farcs$0 \\
\hline
\multicolumn{4}{l}{\footnotesize $^a$ Statistical uncertainty radius. The 90\%\ confidence systematic } \\

\multicolumn{4}{l}{\footnotesize uncertainty is $\sim$56$\arcsec$ for small off axis angles. For larger off-axis } \\
\multicolumn{4}{l}{\footnotesize angles the uncertainty increases to 65--90$\arcsec$ near the edge of the } \\
\multicolumn{4}{l}{\footnotesize FOV (Fiore et al.\ 2000).} \\
\end{tabular}
\end{table}

\begin{table}
\caption{Power-law fit parameters for the LECS/MECS spectra of 1SAX\,J1338.1$-$8030 (X-ray afterglow) and 1SAX\,J1337.6$-$8027}
\begin{tabular}{lccccc}
\hline
Source & N$_{\rm H}$ & Norm$^a$ & photon & F$_{\rm X}$$^b$ & $\chi^2$/dof \\
       & (10$^{21}$\,atoms\,cm$^{-2}$) & & index & (2--10\,keV) & \\
\hline
\multicolumn{6}{l}{\underline{Maximum likelihood method}} \\
1SAX\,J1338.1$-$8030 & 2.1$\pm$0.6 & 0.60$\pm$0.06 & $-$2.03$\pm$0.08 & 1.47$\pm$0.18 & 42/26 \\
1SAX\,J1337.6$-$8027$^c$ & 2.1$^d$ & 0.10$\pm$0.03 & $-$2.1$\pm$0.3 & 0.22$\pm$0.10 & 41/18 \\
\multicolumn{6}{l}{\underline{SPEX method}} \\
1SAX\,J1338.1$-$8030 & 1.7$\pm$0.7 & 0.63$\pm$0.10 & $-$2.12$\pm$0.12 & 1.36$\pm$0.24 &
169/189 \\
1SAX\,J1337.6$-$8027 & `` & 0.09$\pm$0.04 & $-$1.9$\pm$0.4 & 0.27$\pm$0.17 & 
169/189 \\
\multicolumn{6}{l}{\underline{`Canonical' method}} \\
1SAX\,J1338.1$-$8030 & 1.5$\pm$0.6 & 0.62$\pm$0.21 & $-$2.24$\pm$0.13 & 1.13$\pm$0.38 & 86/82 \\
1SAX\,J1337.6$-$8027$^e$ & `` & 0.09 & $-$1.9 & 0.27 & -- \\
\hline
\multicolumn{6}{l}{\footnotesize $^a$ Power-law normalization in 
10$^{-3}$\,photons\,cm$^{-2}$\,s$^{-1}$\,keV$^{-1}$ at 1\,keV.} \\
\multicolumn{6}{l}{\footnotesize $^b$ Unabsorbed X-ray flux (10$^{-12}$\,erg\,s$^{-1}$\,cm$^{-2}$).} \\
\multicolumn{6}{l}{\footnotesize $^c$ MECS only, see text.} \\
\multicolumn{6}{l}{\footnotesize $^d$ Parameter fixed, see text.} \\
\multicolumn{6}{l}{\footnotesize $^e$ Parameters fixed to result from SPEX method, see Appendix.} \\ 
\end{tabular}
\end{table}

\begin{table}
\caption{Power-law fit parameters of MECS spectra of 1SAX\,J1338.1$-$8030 (X-ray afterglow) at 3 time intervals}
\begin{tabular}{cccccc}
\hline
Time$^a$ & Effective exposure & Norm$^b$ & photon & F$_{\rm X}$$^c$ & $\chi^2$/dof \\
(hrs) & ksec & & index & (2--10\,keV) & \\
\hline
8--12.6    &  7.3 & 1.9$\pm$0.3   & $-$2.10$\pm$0.12 & 4.20$\pm$0.75 & 20/13 \\
12.6--21.8 & 12.4 & 1.0$\pm$0.2   & $-$2.13$\pm$0.13 & 2.04$\pm$0.39 &  6/13 \\
21.8--44.3 & 48.2 & 0.37$\pm$0.05 & $-$2.08$\pm$0.12 & 0.85$\pm$0.15 & 21/13 \\
\hline
\multicolumn{6}{l}{\footnotesize $^a$ Time after start of $\gamma$-ray burst trigger.} \\
\multicolumn{6}{l}{\footnotesize $^b$ Power-law normalization in 10$^{-3}$\,photons\,cm$^{-2}$\,s$^{-1}$\,keV$^{-1}$ at 1\,keV.} \\
\multicolumn{6}{l}{\footnotesize $^c$ Unabsorbed X-ray flux (10$^{-12}$\,erg\,s$^{-1}$\,cm$^{-1}$).} \\
\end{tabular}
\end{table}

\newpage

\begin{figure}
\psfig{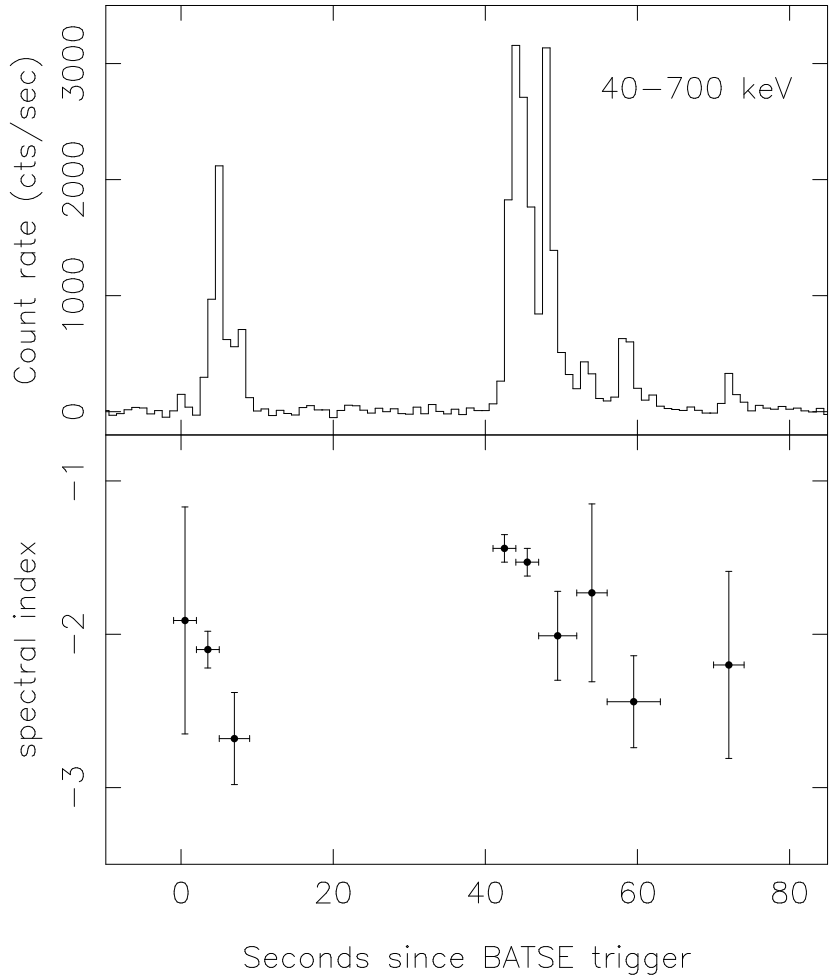}
\caption{Top: 40--700\,keV GRBM light curve of the prompt $\gamma$-ray emission of GRB990510.
Bottom: $\gamma$-ray spectral evolution, as described by the power-law photon index in 
various time windows.}
\end{figure}

\begin{figure}
\psfig{figure=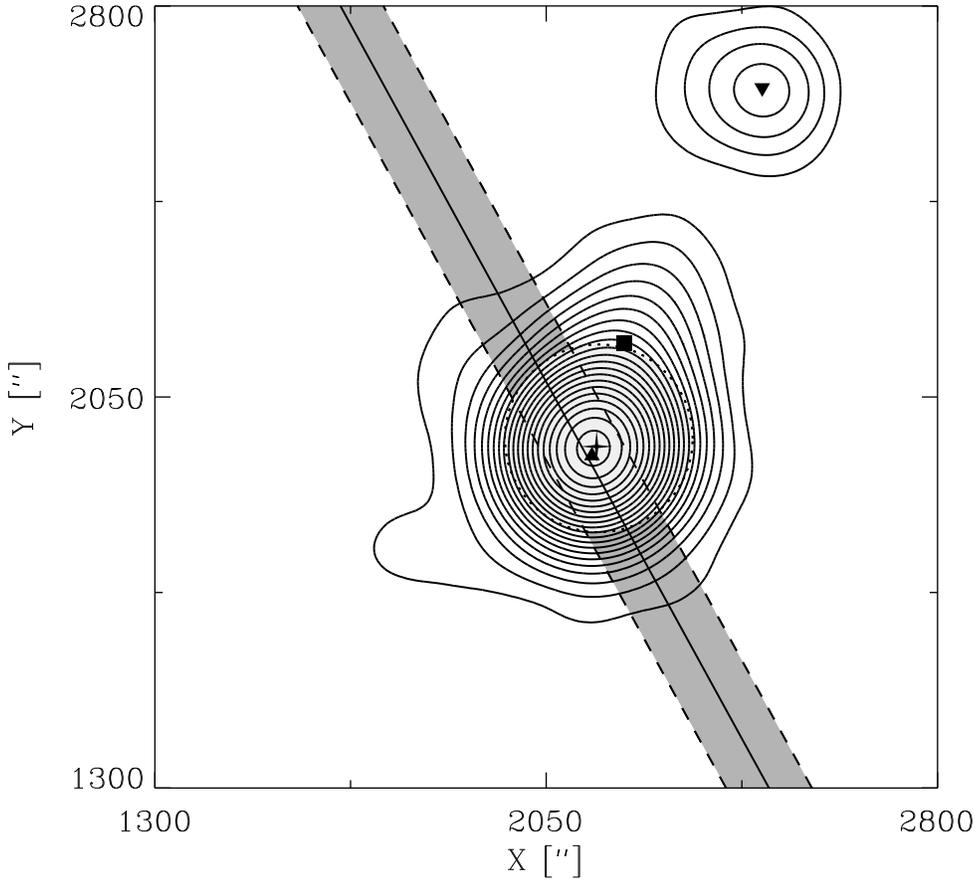,height=13cm}
\caption{Maximum likelihood ratio map (in detector coordinates) of the
sky region of GRB990510 during its X-ray afterglow as observed by the MECS.
North is up and East to the left. The contours start at a $3\sigma$
significance detection level (for 1
degree of freedom)  in steps of $3\sigma$.  The 3 high-significant sources
detected in this field are indicated by a filled triangle
(1SAX\,J1338.1$-$8030 = GRB990510's X-ray afterglow), a filled square
(1SAX\,J1337.6$-$8027) and an inverted filled triangle
(1SAX\,J1336.0$-$8018).  Also indicated as a dark grey shaded band is the
Ulysses/GRBM triangulation annulus bound by its $3\sigma$ half-widths
(Hurley et al.\ 2000), the final WFC error circle as a light grey filled
circle (Dadina et al.\ 1999) and the position of the optical transient as a
star symbol (Vreeswijk et al.\ 1999b).}
\end{figure}

\begin{figure}
\psfig{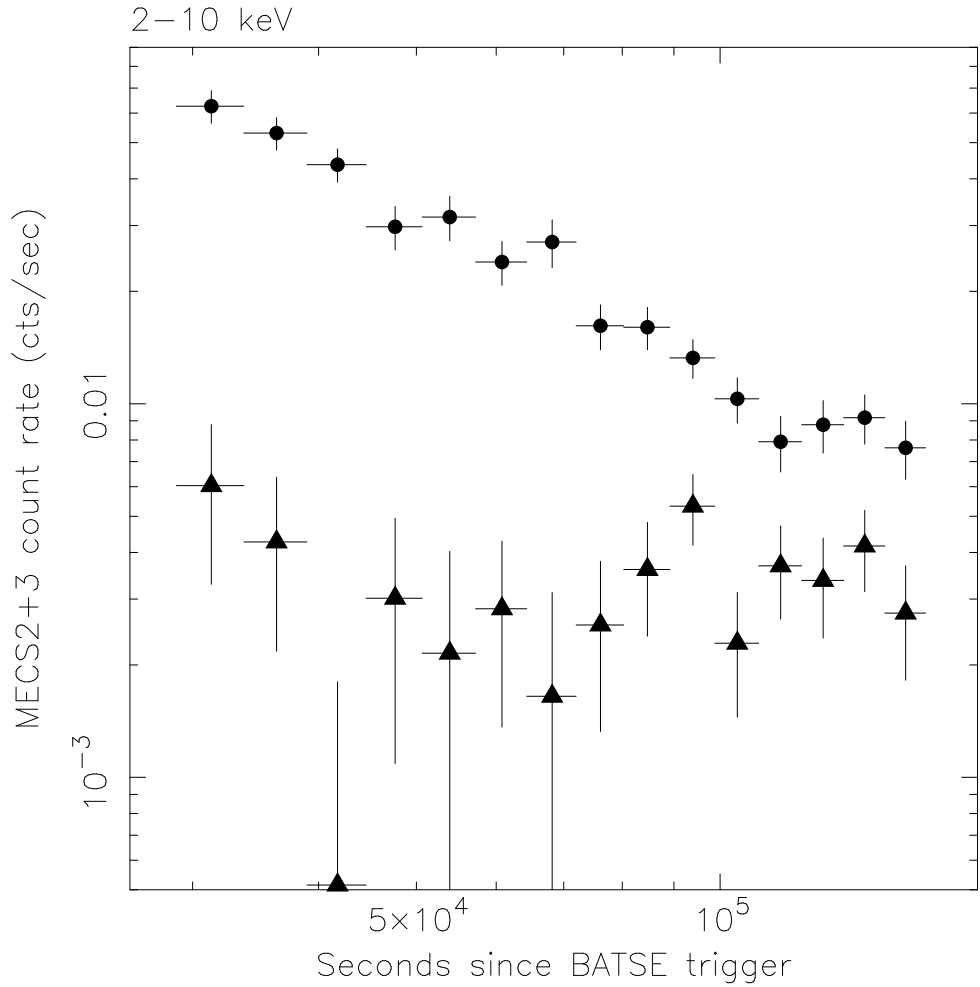}
\caption{2--10\,keV light curves from MECS units 2 and 3 combined for 
1SAX\,J1338.1$-$8030, i.e.\ the X-ray afterglow of GRB990510 (filled circles),
and the nearby ($\sim$3.8\,arcmin) contaminating source 1SAX\,J1337.6$-$8027
(filled triangles). Clearly, the contaminating source is constant throughout the
observation.}
\end{figure}

\begin{figure}
\psfig{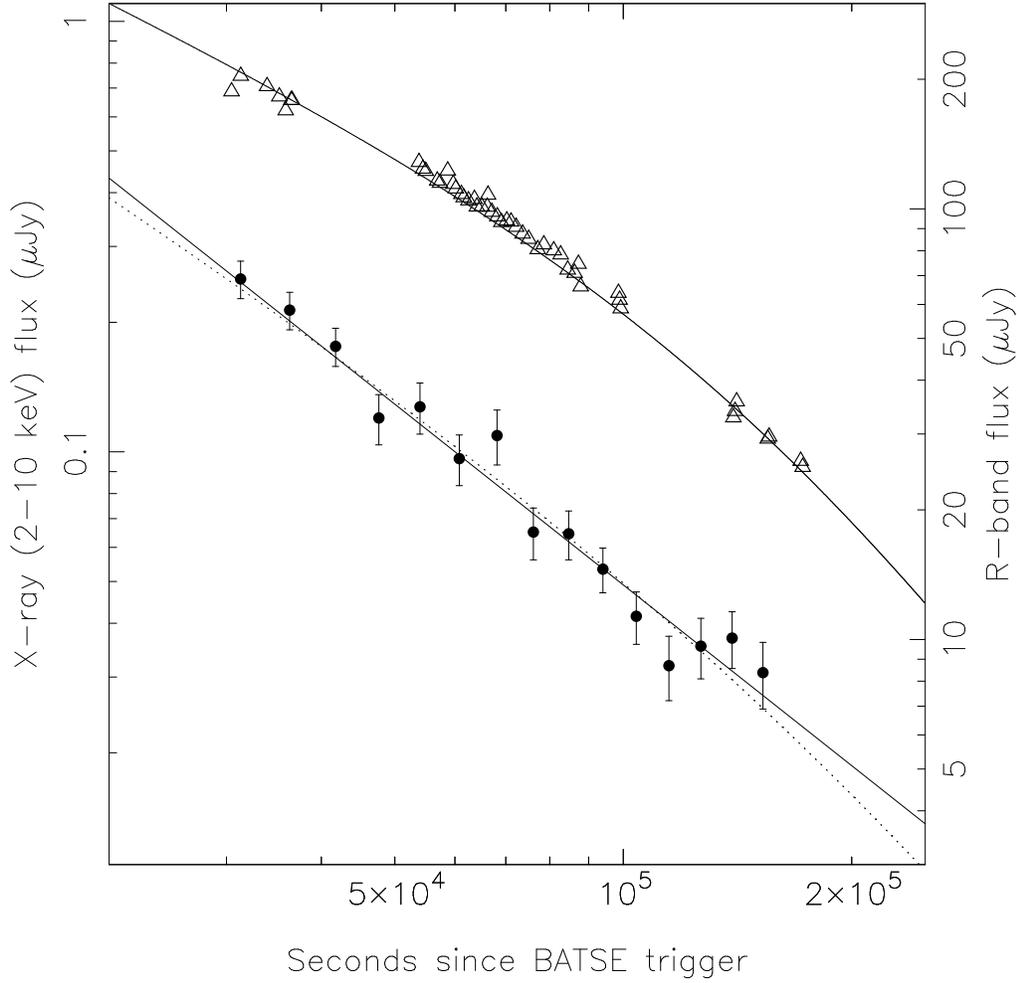}
\caption{X-ray (filled circles) and optical R-band (open triangles) light
curves of the afterglow of GRB990510. The R-band measurements (Stanek et
al.\ 1999, Harrison et al.\ 1999, Vreeswijk et al.\ 1999a, Galama et al.\
1999, Covino et al.\ 1999, Lazzati et al.\ 1999)  have been corrected for
reddening ($A_R=0.54$, Harrison et al.\ 1999) and reduced to the same
photometric system.  For the R-band we also show the
corresponding fit to a
steepening power-law (solid curve) as given in the literature (Harrison et
al.\ 1999,
Stanek et al.\ 1999, Israel et al.\ 1999).  The X-ray light curve can be satisfactorily
described by a single power law (solid line) or a steepening
power
law similar to that describing the R-band data, but with
parameter values appropriate for the X-ray data (dotted line; see text).}
\end{figure}

\end{document}